\documentclass[prl,superscriptaddress,twocolumn,showpacs,preprintnumbers,%
amsmath,amssymb]{revtex4}

\usepackage{graphicx}
\usepackage{dcolumn}
\usepackage{bm}

\begin{document}   
\title{Self-localization of composite spin-lattice polarons}

\author{Peter Prelov\v sek}
\affiliation{Max-Planck-Institut f\"ur Festk\"orperforschung,
             Heisenbergstrasse 1, D-70569 Stuttgart, Germany} 
\affiliation{J. Stefan Institute, SI-1000 Ljubljana, Slovenia and }
\affiliation{Faculty of Mathematics and Physics, University
             of Ljubljana, SI-1000 Ljubljana, Slovenia }
\author{Roland Zeyher}
\affiliation{Max-Planck-Institut f\"ur Festk\"orperforschung,
             Heisenbergstrasse 1, D-70569 Stuttgart, Germany}
\author{Peter Horsch}
\affiliation{Max-Planck-Institut f\"ur Festk\"orperforschung,
             Heisenbergstrasse 1, D-70569 Stuttgart, Germany}

\date{\today}

\begin{abstract}
Self-localization of holes in the Holstein $t$-$J$ model 
is studied  in the adiabatic limit using
exact diagonalization and the retraceable path approximation. It is shown
that the critical electron-phonon coupling $\lambda_c$ decreases with
increasing $J$ and that this behavior is determined mainly by the
incoherent rather than by the coherent motion of the hole. 
The obtained spin correlation functions in the localized region can be 
understood within a percolation picture where antiferromagnetic
order can persist up to a substantial hole doping. These results restrict 
the possibility of self-localization of holes in lightly doped cuprates.
\end{abstract}


\maketitle


The interplay of strong electronic correlations and electron-phonon (EP) 
interactions in the formation of dressed quasi-particles
is one of central puzzles of high-$T_c$ superconductors.
Recent ARPES experiments in undoped cuprates were interpreted 
in terms of strong EP coupling giving rise to self-localization of holes
\cite{Roe05,She05,Mis04}. Similar effects
can be expected in the manganites in the colossal
magnetoresistance regime, where polarons are not only dressed by
spin and lattice excitations but also polarize the orbitals in the 
neighborhood of the holes. The properties of composite polarons are subtle
and not well understood, as the different mechanisms involved may support
or compete with each other \cite{Zho92,Feh93,Roe93,Feh95,Gre96,Ram92}.

Electronic motion in weakly doped Mott insulators is determined by the 
constraint of no double occupancy of sites which renders the motion of 
holes predominantly incoherent with an energy scale $t$, the bare 
hopping energy of electrons. In addition, the antiferromagnetic (AF) 
exchange interaction $J$ allows for spin flips leading to
coherent hole motion 
at the bottom of the band
with a dispersion characterized by $J$.
In the presence of strong EP coupling the spin polaron
transforms into a spin-lattice polaron. The formation of this 
composite polaron affects in general both the coherent translational
as well as the internal incoherent motion.
In recent studies of Holstein $t$-$J$ models appropriate for cuprates,
it has been found that
the effect of EP-interaction on spin-polarons is strongly enhanced
as compared to polarons in uncorrelated systems.
In particular, the critical EP-coupling $\lambda_c$ for self-localization of 
composite polarons is significantly reduced
\cite{Roe05,Mis04,Zho92,Feh93,Roe93,Feh95,Gre96}, and in the
regime $\lambda<\lambda_c$ the coherent polaron mass is strongly enhanced 
as compared to the uncorrelated case \cite{Ram92,Mis04}. 

In spite of several studies there are important open questions, which we 
address in the following:
(i) Is the critical EP coupling $\lambda_c$ for self-localization determined
by the coherent bandwidth or rather by the incoherent hole motion?
(ii) What is the dependence of $\lambda_c$ on $J$?
(iii) Furthermore, is the observed low doping concentration
$x \sim 0.02$ for the destruction of AF long-range
order in cuprates compatible with self-localized
polarons? 

To answer the above questions we investigate the Holstein $t$-$J$ model
in the adiabatic limit where the kinetic energy of the lattice can be
neglected:
\begin{eqnarray}
H&=&
-\sum_{<ij>\sigma}t_{ij}(\tilde{c}^+_{i\sigma}\tilde{c}_{j\sigma}+ h.c)
+ J\sum_{<ij>}({\bf S}_i{\bf S}_j-\frac{1}{4}n_i n_j) \nonumber \\
& &-g\sum_i u_i n_i^h + \frac{K}{2}\sum_i u_i^2.
\label{H}
\end{eqnarray}
The first two terms in Eq.(\ref{H}) represent the Hamiltonian
$H_{t-J}$ of the $t$-$J$ model.
The transfer matrix elements $t_{ij}$ include
both nearest and next-nearest neighbor hoppings $t$ and $t'$,
respectively. The third term describes the lattice potential  
$H_{ep}$ proportional to the
EP coupling constant $g$, the local displacement fields $u_i$, and the
hole density operator $n_i^h=1-n_i=1-\sum_{\sigma}
\tilde{c}^+_{i\sigma}\tilde{c}_{i\sigma}$.
In high-T$_c$ cuprates this term reflects the breathing motion of oxygen
ions around the hole and the subsequent change of the Zhang-Rice singlet 
energy due to the local lattice distortion.
The last term is the elastic energy $H_{ph}$ 
with force constant $K$. In the adiabatic limit displacements $u_i$
can be treated classically and are in the equilibirum 
determined by $u_i=(g/K)<n_i^h>$.
At vanishing and weak EP-coupling $H$ exhibits coherent (delocalized)
quasi-particles with bandwidth $\sim J$. Nevertheless, we will show in
the following that many features of $H$ can be very well represented
by the simpler Holstein $t$-$J_z$ model in the retraceable path approximation
(rpa), where the motion of carriers is entirely incoherent.
It is convenient to express results in terms of the 
dimensionless EP coupling parameter $\lambda=g^2/8 K t$. For noninteracting
electrons we have then $\lambda_c \sim 1$. In the following we also put 
$t=K=1$.

The Hamiltonian Eq.(\ref{H}) is solved by exact diagonalization (ED)
of small planar systems with a square lattice. 
The determination of
the critical EP coupling is straightforward, namely, it separates ground
states with homogeneous and inhomogeneous lattice dispacements $u_i$
corresponding to a delocalized and a self-localized polaron solution, 
respectively. Since the
itinerant ground state for a hole in the $t$-$J$ model is at ${\bf
k}_0 \sim (\pi/2,\pi/2)$, the ground state is degenerate. This degeneracy
leads in general to inhomogeneous $u_i$ which spoil the
interpretation \cite{Feh93,Roe93}. We avoid this problem by choosing
twisted boundary conditions which lift the degeneracy. For fixed $u_i$
we find then the ground state using ED for systems with $N=18$ and 20
sites, whereby the equilibrium is reached by the iteration of the
self-consistency relation $u_i= (g/K)<n_i^h>$.

The $t$-$J_z$ Holstein model is solved within rpa,
which excludes loop motion and thus neglects a very small coherent bandwidth
that arises at small $J_z$ from loop trajectories \cite{Dag94}.
Using a Bethe lattice the expectation value 
of the
electronic energy, namely, $\langle H_{t-J_z}\rangle +
\langle H_{ep} \rangle$, is given by the lowest energy
eigenvalue $\epsilon_0$ of the symmetric matrix $H_{ll'}$ with the elements
\begin{equation}
H_{00} = J_z-gu_0,\;\; H_{ll}=({3\over 2} +l)J_z-gu_l \;\;\mbox{for}\;\;
l=1,2..,
\label{H00}
\end{equation}
\begin{equation}
H_{01}=-2t,\;\; H_{ll+1} =-\sqrt{3}t \;\;\mbox{for}\;\; l=0,1,..
\label{H01}
\end{equation}
Here $l=0,1,2..$ denotes the shell of l-th nearest neighbors consisting of
$N_l$ ions with $N_0=1$ and $N_l=4\cdot 3^{l-1}$ for $l=1,2,..$, and
$u_l$ is the displacement of one of the equivalent ions in the shell $l$.
The expectation value of the density at the ion $i$, $\langle
n_i^h \rangle$, is equal to $|e(0,l)|^2/N_l$, where $e(0,l)$ is the
normalized eigenvector belonging to the lowest eigenvalue $\epsilon_0$
and $l$ is the shell index of the ion $i$. Finally, the total energy
$\epsilon_0 + \langle H_{ph} \rangle$ is minimized with respect
to the displacements $\{u_l\}$.   

\begin{figure}
\includegraphics[width=6cm]{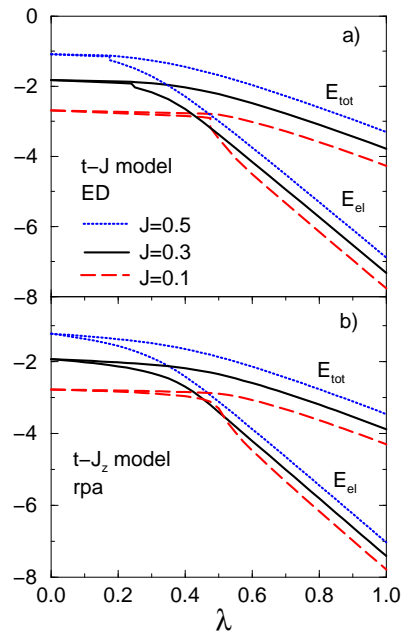}
\caption{\label{fig:1}
(color online)
Electronic  and total energies $E_{el}$ and $E_{tot}$, respectively, as a 
function of $\lambda$ for (a) the Holstein $t$-$J$ model using ED and (b)
the Holstein $t$-$J_z$ model using rpa. The inset explaining the curves in (a)
also holds for (b).
}
\vspace{-0.7cm}
\end{figure}

Fig. \ref{fig:1}(a) shows the electronic energy $E_{el} = 
\langle H_{t-J} \rangle +\langle H_{ep} \rangle$
and the total energy $E_{tot} = E_{el} + \langle H_{ph} \rangle$ as a
function of $\lambda$ for the Holstein $t-J$ model 
and $J=0.1, 0.3$ and $0.5$. The results were obtained by ED for one single
hole and clusters with $N=20$ sites. For $\lambda \gg 1$ the hole is
completely localized yielding $E_{el} = 2 E_{tot}
\sim -8t \lambda$. In the limit $\lambda \rightarrow 0$ the well-known
single-hole energies $E^0_{el}$ of the $t-J$ model are reproduced.
For $J \rightarrow 0$ the energy is close to the rpa result
$E^0_{el} \sim -2\sqrt{3}t$ (Note that we avoid in the ED studies very
small values for $J$ where the ground state is Nagaoka-type ferromagnetic 
with $E^0_{el} \sim -4t$). For a finite $J$  
$E_{el}$ increases because the itinerant hole weakens the antiferromagnetic
bonds. Using ED results for spin correlation functions 
this increase is about $4.72 J$ which agrees well with Fig.\ref{fig:1}(a).
For $\lambda < \lambda_c$ the homogenous solution with $E^0_{el}$ is
stable (the small slope appearing in  Fig.\ref{fig:1}(a) is a finite size
effect since $u_i \sim 1/N$). For $\lambda > \lambda_c$ the localized
solution has the lowest total energy approaching the linear 
behavior at large $\lambda$. Due to the first-order transition between
itinerant and localized solutions the ED data exhibit a small jump in $E_{el}$
and a change in slope in $E_{tot}$ at $\lambda_c$. 

Fig. \ref{fig:1}(b) shows the same quantities as in Fig. \ref{fig:1}(a) but 
calculated for the Holstein $t$-$J_z$ model using the rpa.
The curves are nearly identical with those of Fig. \ref{fig:1}(a). 
The main difference
is the absence of a well defined transition, at least in the
cases $J=0.3$ and 0.5. For $J=0.1$ $E_{tot}$ and, to a lesser degree,
$E_{el}$ are practically constant
up to $\lambda \sim 0.4$. While for small $J$ $E_{tot}$ decreases gently
with increasing $\lambda$ $E_{el}$ shows a sudden decrease 
at $\lambda \sim 0.5$ before reaching
its asymptotic linear behavior. Similar features are seen in the curve
for $J=0.1$ in Fig. \ref{fig:1}(a) obtained by ED. In the limit
$J \rightarrow 0$ $E_{tot}$ and $E_{el}$ are identical and equal to 
$-\sqrt{12}t$ up to
a critical coupling $\lambda_c^{(0)} \sim 0.580$. For $\lambda >
\lambda_c^{(0)}$ there exist two solutions of the extremal equation.
One is the homogenous solution where $E_{tot}$ and $E_{el}$ are identical and
independent of $\lambda$. The second solution describes a localized polaron
where $E_{tot}$ exhibits an upward jump by about 0.4 and
$E_{el}$ a downward jump by about 0.6 at $\lambda_c^{(0)}$. This means that
the localized solution is unstable in the interval $[\lambda_c^{(0)},
\lambda_c]$, where $E_{tot}$ of the localized solution crosses $E_{tot}$ 
of the homogenous solution at $\lambda_c \sim 0.662$.   
With increasing $J$ the jumps are replaced by cross-overs with decreasing
absolute changes until for $J >0.06$ there exists only one solution of the 
extremal equation describing a crossover from an extended
to a localized polaron. 

\begin{figure}
\includegraphics[width=8cm]{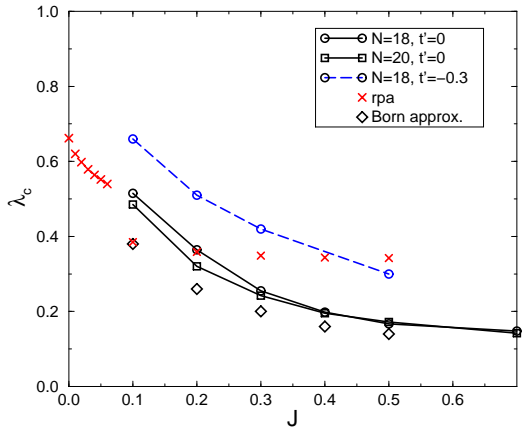}
\vspace{-0.3cm}
\caption{\label{fig:2}
(color online)
Dependence of $\lambda_c$ on $J$ comparing results from ED, rpa, and Born 
approximation.
}
\vspace{-0.5cm}
\end{figure}

Fig. \ref{fig:2} shows ED data (circles and squares) 
for $\lambda_c$ as a function of $J$. 
The dashed curve represents
data for $t'=-0.3$, i.e., a case with a larger and more dispersive
coherent part than for $t'=0$. As a result, the dashed curve lies above
the ED data for $t'=0$, showing, that decreasing the mass $M$ for the coherent
motion leads to an increase in $\lambda_c$ as expected. Increasing $J$ also
decreases $M$ but Fig. \ref{fig:2} implies that in this case $\lambda_c$ 
decreases and does not increase. 
This is unexpected, since an interpretation in terms of an opposite trend 
$\lambda_c\sim J$
has been given in a study of the Holstein $t$-$J$ model \cite{Feh95}.

In order to understand the dependence of $\lambda_c$ on $J$ we
have estimated $\lambda_c$ using the rpa and the Born approximation.
For $J < 0.06$ the rpa leads to a first-order transition between a 
localized and delocalized ground state, the resulting $\lambda_c$'s are shown 
as crosses in Fig. \ref{fig:2}. 
In order to find a sharp transition 
at larger values of $J$ the energy of the homogeneous solution has to be
compared with the localized one.
An estimate for the loss of coherent kinetic energy in the
localized state is $\Delta E_{kin} =
\bar{\epsilon} -\epsilon({\bf k}_0) \sim 0.65J$ \cite{Dag94}
where $\epsilon({\bf k})$
is the quasiparticle dispersion and $\bar{\epsilon}$ its average value.
$\lambda_c$ then follows from the rpa result via the relation
$\Delta E_{kin} = E_{tot}(0)- E_{tot}(\lambda_c)$. The resulting
values (crosses for $J \geq 0.1$
in Fig. \ref{fig:2}) give the right trend but cannot reproduce quantitatively
the strong decrease of $\lambda_c$ with $J$ of the ED results.
Modifying the Born approximation\cite{Ram92} to take inhomogenous local
potentials into account while retaining the homogenous self-energy
we have determined the smallest value of $\lambda$ where a localized 
solution exists and identified this value with $\lambda_c$. 
The resulting diamonds in Fig. \ref{fig:2} are rather near to the ED data.

The above results may be interpreted in simple physical terms as 
follows: At $J=0$ the hole motion is entirely
incoherent and determined by the energy scale $t$.
The rpa solution shows in this region a sharp transition at $\lambda_c$
from a very localized polaron to a very extended object. The obtained
critical value $\lambda_c \sim 0.6$ is somewhat reduced compared to
the free case $\sim 0.85$. For larger $J$ the transition in rpa evolves into 
a pronounced crossover in $E_{tot}(\lambda)$ as is evident from Fig. 2b),
where the crossover value $\lambda^\ast$ moves to lower $\lambda$'s
due to the reduced polaron radius. Although the sharp transition
finally involves $\Delta E_{kin} \propto J$, $\lambda_c$ is effectively
governed by the incoherent solution and its $\lambda^\ast$.
The decrease of $\lambda_c$ with $J$ reflects
a smooth transition from the large energy scale $t$ characteristic for
the incoherent motion to the scale $J$ relevant for the coherent part. 
This picture may also explain ED data showing that $\lambda_c$ actually 
increases with $J$ at large $J \sim t$ where the incoherent part is small
compared to the coherent one. 
\begin{figure}
\includegraphics[width=7.2cm]{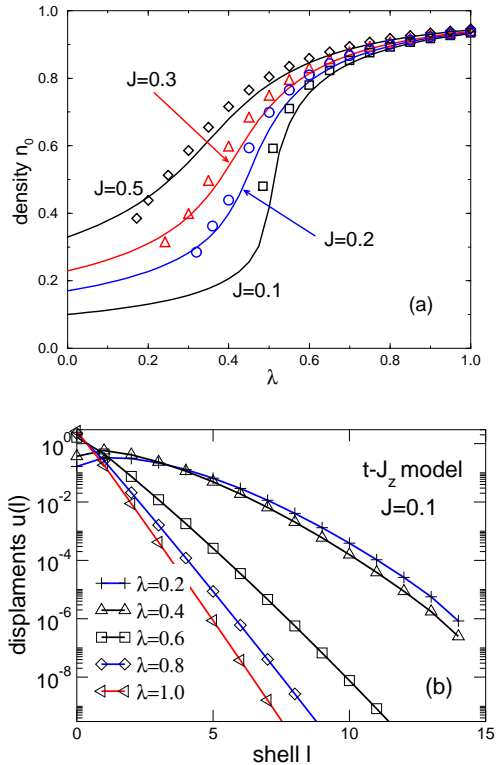}
\vspace*{0mm}
\centerline{\includegraphics[angle=0,width=6.5cm]{figpolaron3b.eps}}
\caption{\label{fig:3}
(color online)
(a) Density $n_0$ in the center of the polaron as a function of 
$\lambda$.
Results obtained by ED for the $t$-$J$ model (symbols) are compared with
rpa results for the $t$-$J_z$ model (lines).  
(b)  Log-linear plot of displacement fields $u_l$ 
of the Holstein t-J$_z$ model versus distance $l$ showing the exponential 
decay of hole distribution in the self-localized regime.
}
\vspace{-0.5cm}
\end{figure}
Fig. \ref{fig:3}(a) shows the hole density at the center of the localized 
state, denoted by $n_0$, as a function of $\lambda$ for different values of 
$J$. The symbols are results from the ED of clusters with
$N=20$ sites in the localized region, the lines correspond to the rpa. 
$n_0=1$ corresponds to a completely localized and
$n_0 =1/N$ to an extended state.
The agreement between both results in Fig. \ref{fig:3}(a) is 
excellent, showing, that the degree of localization of the polaron
is to a very good approximation determined by the incoherent part
of the hole motion described well by the rpa. In the limit $J \rightarrow 0$
the rpa yields a curve for $n_0$ which is zero for $\lambda < \lambda_c^{(0)}$,
jumps to about $0.65$ at $\lambda_c^{(0)}$, and then smoothly approaches
the $J=0.1$ curve. The crossover value $\lambda^\ast$ is clearly visible in 
Fig. \ref{fig:3}(a) and can be identified as the inflection point in the 
$n_0(\lambda)$ curves. 
Fig. \ref{fig:3}(b) illustrates the qualitative change
of the displacements $u_l$ at $\lambda_c$, i.e., for $\lambda > \lambda_c$
$u_l$ decays exponentially with distance $l$.

An important consequence for magnetism follows from 
self-localization of holes in weakly doped high-T$_c$ superconductors.
It is well known that AF long-range order vanishes in hole-doped cuprates
at quite low doping concentration $x_{AF}\sim 0.02-0.04$.
This small value is due to the incoherent motion of holes which destroys
the AF order; on the other hand a percolative 
model based on static holes and broken AF bonds leads to much larger critical 
concentrations $x_{AF}\sim 0.5$.
Self-localization at strong EP coupling $\lambda>\lambda_c$ would hinder 
the incoherent motion of holes and preserve the antiferromagnetic order.
\begin{figure}
\includegraphics[width=7cm]{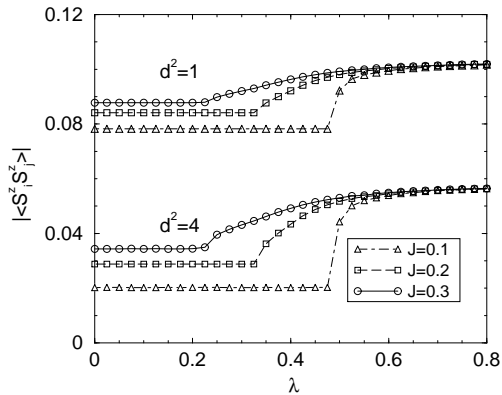}
\vspace{-0.3cm}
\caption{\label{fig:4}
Averaged spin correlation function $<S_i^z S_j^z>$ for the Holstein 
$t$-$J$ model using $ED$, 1 hole, and 18 sites. $d^2=1,4$ denote 
first- and third-nearest neighbors.
}
\vspace{-0.5cm}
\end{figure}

The strong enhancement of AF correlations in the self-localized regime
$\lambda >\lambda_c$ is displayed in 
Fig. \ref{fig:4} which shows ED data for the first ($d^2=1$) and third 
($d^2=4$) nearest
neighbor spin correlation function (SCF) of the $t$-$J$ model.
For the undoped Heisenberg antiferromagnet the 
corresponding SCF
values are -0.116 and 0.067, correspondingly. For $\lambda >> \lambda_c$  
the hole is totally localized and one expects a decrease of these values
due to the four broken bonds yielding a reduction factor of 16/18 for 18 sites.
The resulting values -0.103 and 0.059 agree quite well with the
numerical data at large $\lambda$ in Fig. \ref{fig:4}, suggesting
the applicability of a simple percolation picture at large $\lambda$. 
The figure also illustrates that for a given type of neighbors the SCF
become independent of $J$ well above $\lambda_c$ which reflects
the fact that the localizing EP coupling and not the string potential
$\sim J$, Eq.(\ref{H00}), determines the extension of the polaron. 
The slight decrease of SCF's with decreasing $\lambda$
in the localized region can be understood in the rpa. With decreasing 
$\lambda$ the hole explores more and more neighboring sites producing hereby
an increasing number of broken bonds which reduce the SCF. Such a
picture reproduces for $\lambda > 0.6$ quantitatively the ED 
data for SCF in the Ising, and to a good degree, also in the Heisenberg case.
This suggests
that for the considered doping $x \sim 0.05$ long-range AF order persists
throughout the localized region and that immobile holes 
reduce mainly the magnitude of the SCF's but not their spatial decay.
 
In conclusion, we have shown that the critical EP coupling $\lambda_c$
for self-localization in the Holstein $t$-$J$ model decreases with $J$ and
that the crossover between basically incoherent (coherent) motion of the
hole at small (large) $J$ is 
responsible for this decrease. Exact diagonalization results for 
$\lambda_c$, the density and displacement distribution as well as 
spin correlation functions have been obtained and successfully interpreted
in the localized regime within the retraceable path approximation. 
Our results suggest that for dopings $x \sim 0.05$ self-localized holes
cannot suppress sufficiently antiferromagnetic correlations to explain the
observed absence of long-range order in the cuprates at this doping level.
We therefore believe that holes in lightly doped cuprates are not 
self-localized,
and that their EP coupling is below the critical value $\lambda_c \sim 0.25$
for $J=0.3$.   
Our conclusions are consistent with the observation of a sharp quasi-particle 
peak in ARPES experiments of $3\;\%$ doped LaSrCuO$_4$\cite{Yosh03}, and also
agree with the explanation of transport and optical
data in lightly doped La$_2$CuO$_4$ in terms of holes bound to 
impurities at low but delocalized at high temperatures \cite{Fal93,Che95}.

Useful discussions with O. Gunnarsson are acknowledged.

\end{document}